\def\papertitle{Data Augmentation for Instrument Classification Robust to Audio Effects}
\def\paperauthorA{Ant\'{o}nio Ramires}
\def\paperauthorB{Xavier Serra}
\def\paperauthorC{}
\def\paperauthorD{}

\documentclass[twoside,a4paper]{article}
\usepackage{dafx_19}
\usepackage{amsmath,amssymb,amsfonts,amsthm}
\usepackage{euscript}
\usepackage[latin1]{inputenc}
\usepackage[T1]{fontenc}
\usepackage{ifpdf}
\usepackage{booktabs}

\usepackage[english]{babel}
\usepackage{caption}
\usepackage{subfig} 
\usepackage{color}
\usepackage{url}
\usepackage{multirow}

\ninept

\usepackage{cite}
\usepackage{times}

\newif\ifpdf
\ifx\pdfoutput\relax
\else
 \ifcase\pdfoutput
  \pdffalse
 \else
  \pdftrue
\fi

\ifpdf 
 \usepackage[pdftex,
 pdftitle={\papertitle},
 pdfauthor={\paperauthorA, \paperauthorB, \paperauthorC, \paperauthorD},
 colorlinks=false, 
 bookmarksnumbered, 
 pdfstartview=XYZ 
 ]{hyperref}
 \usepackage[pdftex]{graphicx}
\else 
 \usepackage[dvips]{epsfig,graphicx}
 \usepackage[dvips,
 colorlinks=false, 
 bookmarksnumbered, 
 pdfstartview=XYZ 
 ]{hyperref}
\fi
 \usepackage[hypcap=true]{caption}
\title{\papertitle}

\twoaffiliations{
\paperauthorA }
{\href{https://www.upf.edu/web/mtg}{Music Technology Group} \\ Universitat Pompeu Fabra \\ Barcelona, Spain \\ {\tt \href{mailto:antonio.ramires@upf.edu}{antonio.ramires@upf.edu}}}
{\paperauthorB }
{\href{https://www.upf.edu/web/mtg}{Music Technology Group} \\ Universitat Pompeu Fabra \\ Barcelona, Spain \\ {\tt \href{mailto:antonio.ramires@upf.edu}{xavier.serra@upf.edu}}}

\begin{document}
\ifpdf 
 \DeclareGraphicsExtensions{.png,.jpg,.pdf}
\else 
 \DeclareGraphicsExtensions{.eps}
\fi

\maketitle
\sloppy
\begin{abstract}

Reusing recorded sounds (sampling) is a key component in Electronic Music Production (EMP), which has been present since its early days and is at the core of genres like hip-hop or jungle. Commercial and non-commercial services allow users to obtain collections of sounds (sample packs) to reuse in their compositions. Automatic classification of one-shot instrumental sounds allows automatically categorising the sounds contained in these collections, allowing easier navigation and better characterisation.

Automatic instrument classification has mostly targeted the classification of unprocessed isolated instrumental sounds or detecting predominant instruments in mixed music tracks. For this classification to be useful in audio databases for EMP, it has to be robust to the audio effects applied to unprocessed sounds. 

In this paper we evaluate how a state of the art model trained with a large dataset of one-shot instrumental sounds performs when classifying instruments processed with audio effects. In order to evaluate the robustness of the model, we use data augmentation with audio effects and evaluate how each effect influences the classification accuracy.
\end{abstract}

\section{Introduction}
\label{sec:intro}
The repurposing of audio material, also known as sampling, has been a key component in Electronic Music Production (EMP) since its early days and became a practice which had a major influence in a large variety of musical genres. The availability of software such as Digital Audio Workstations, together with the audio sharing possibilities offered with the internet and cloud storage technologies, led to a variety of online audio sharing or sample library platforms. In order to allow for easier sample navigation, commercial databases such as sounds.com\footnote{https://sounds.com/} or Loopcloud\footnote{https://www.loopcloud.net/} rely on expert annotation to classify and characterise the content they provide. In the case of collaborative databases such as Freesound \cite{FS} the navigation and characterisation of the sounds is based on unrestricted textual descriptions and tags of the sounds provided by users. This leads to a search based on noisy labels which different members use to characterise the same type of sounds.

Automatically classifying one-shot instrumental sounds in unstructured large audio databases provides an intuitive way of navigating them, and a better characterisation the sounds contained. For databases where the annotation of the sounds is done manually, it can be a way to simplify the job of the annotator, by providing suggested annotations or, if the system is reliable enough, only presenting sounds with low classification confidence. 

The automatic classification of one-shot instrumental sounds remain an open research topic for music information retrieval (MIR). While the research on this field has been mostly performed on clean and unprocessed sounds, the sounds provided by EMP databases may also contain ``production-ready'' sounds, with audio effects applied on them. Therefore, in order for this automatic classification to be reliable for EMP sample databases, it has to be robust to the types of audio effects applied to these instruments. In our study, we evaluate the robustness of a state of the art automatic classification method for sounds with audio effects, and analyse how data augmentation can be used to improve classification accuracy.

\section{Related Work}

Automatic instrument classification can be split into two related tasks with a similar goal. 
The first is the identification of instruments in single instrument recordings (which can be isolated or overlapping notes) while the second is the recognition of the predominant instrument in a mixture of sounds. A thorough description of this task and an overview of the early methodologies used is presented in \cite{automaticreview}. These early approaches used two modules for classification, one for extracting and selecting handcrafted features (e.g. Mel Frequency Cepstral Coefficients, spectral centroid, roll-off, and flux) and another for classification (e.g. k-nearest neighbours, support vector machines or hidden Markov models). Datasets used for the evaluation and training of these algorithms included RWC \cite{goto2002rwc} or the University of Iowa Musical Instrument Samples\footnote{\url{http://theremin.music.uiowa.edu/MIS.html}}. While these datasets are small (RWC has $50$ instruments) they proved to be good for classification using handcrafted features. New datasets such as IRMAS \cite{bosch2012comparison} for predominant instrument classification and GoodSounds \cite{goodsounds} with single instrument recordings have been created and provided sufficient data for deep learning approaches to be able to surpass more traditional machine learning approaches. A review of the evolution of traditional machine learning and deep learning approaches for instrument classification is presented in \cite{venkatesh_shenoy_kadandale_2018_1468051}. While the performance of traditional machine learning methods rely on developing handcrafted features, deep learning methods learn high-level representations from data using a general-purpose learning procedure, eliminating the need of expert feature extraction \cite{lecun2015deep}. However, the success of these approaches is highly dependent on both the type and amount of data they are provided \cite{perez2017effectiveness}. 

Recent work has shown the effectiveness of using Convolutional Neural Networks (CNNs) for instrument classification \cite{pons2017timbre, classificationHan, li2015automatic, park2015musical}. CNNs can be seen as trainable feature extractors, where kernels (or filters) with trainable parameters are convolved over an input, being able to capture local spatial and temporal characteristics. This architecture has been applied with great success to the detection, segmentation and recognition of objects and regions in images \cite{lecun2015deep}. In the audio domain, when raw audio or spectograms are given, CNNs are able to learn and identify local spectro-temporal patterns relevant to the task to which they are applied. When utilized for MIR tasks, CNNs have outperformed previous state of the art approaches for various tasks \cite{choi,classificationHan}. For automatic instrument classification, the state of the art approaches use CNNs trained on different representations of the input, such as raw audio \cite{li2015automatic}, spectograms together with multiresolution recurrence plots \cite{park2015musical} and log mel-frequency spectograms \cite{classificationHan, pons2017timbre}. In \cite{pons2017timbre}, CNNs were tailored towards learning timbre representations in log mel-frequency spectograms through the use of vertical filters instead of the commonly used square filters. For instrument classification, this approach displays a close to the state of the art \cite{classificationHan} accuracy on the IRMAS dataset \cite{bosch2012comparison}, while reducing the number of trainable parameters by approximately 23 times, on the single-layer proposed model.

Within the context of NSynth \cite{nsynth2017}, a new high-quality dataset of one shot instrumental notes was presented, largely surpassing the size of the previous datasets, containing $305979$ musical notes with unique pitch, timbre and envelope. The sounds were collected from $1006$ instruments from commercial sample libraries and are annotated based on their source (acoustic, electronic or synthetic), instrument family and sonic qualities. The instrument families used in the annotation are bass, brass, flute, guitar, keyboard, mallet, organ, reed, string, synth lead and vocal. The dataset is available online\footnote{\url{https://magenta.tensorflow.org/datasets/nsynth}} and provides a good basis for training and evaluating one shot instrumental sound classifiers. This dataset is already split in training, validation and test set, where the instruments present in the training set do not overlap with the ones present in validation and test sets. However, to the best of our knowledge, no methods for instrument classification have so far been evaluated on this dataset.

In order to increase the generalisation of a model further than the data provided to it, one possible approach is to use data augmentation. This approach can be described as applying deformations to a collection of training samples, in a way that the correct labels can still be deduced \cite{mcfee2015software}. In computer vision, transforming images by cropping, rotation, reflection or scaling are commonly used techniques for data augmentation. In the audio domain, an intuitive and practical transformation is applying audio effects to the original training audio files. Transformations such as time-stretching, pitch-shifting, dynamic range compression and adding background noise have been applied with success to environmental sound classification, for overcoming the data scarcity problems \cite{salamon2017deep}. In \cite{ko2017study}, artificial reverberation was applied to speech recordings, so as to create a speech recognition system robust to reverberant speech. For instrument recognition, the same set of effects used in \cite{salamon2017deep} was applied with success in \cite{mcfee2015software}. We believe that the use of audio effects typically used in EMP such as echo, reverb, chorus, saturation, heavy distortion or flanger can lead to a useful augmentation, as well as to an increase in robustness in instrument classification scenarios where the instrument recordings have these effects applied. 

\section{Methodology}

In our study we will conduct two experiments. First, we will try to understand how augmenting a dataset with specific effects can improve instrument classification and secondly, we will see if this augmentation can improve the robustness of a model to the selected effect.

To investigate this, we process the training, validation and test sets of the NSynth \cite{nsynth2017} dataset with audio effects. A state of the art deep learning architecture for instrument classification \cite{pons2017timbre} is then trained with the original training set, and appended with each of the augmented datasets for each effect. We use the model trained with the original training set as a baseline and compare how the models trained with augmented versions perform on the original test and on the augmented versions of it for each effect. The code for the experiments and evaluation is available in a public GitHub repository\footnote{\url{https://github.com/aframires/instrument-classifier/}}.

\subsection{Data Augmentation and Pre-Processing}

The audio effects for the augmentation were applied directly to the audio files present in the training, validation splits of the NSynth dataset \cite{nsynth2017}. For the augmentation procedure, we used a pitch-shifting effect present in the LibROSA\footnote{\url{https://librosa.github.io/librosa/}} library and audio effects in the form of VST audio plugins. For the augmentation which used audio plugins, the effects were applied directly to the audio signals using the Mrs. Watson\footnote{\url{https://github.com/teragonaudio/MrsWatson}} command-line audio plugin host. This command line tool was designed for automating audio processing tasks and allows the loading of an input sound file, processing it using a VST audio effect and saving the processed sound. In order to maintain transparency and reproducibility of this study only VST plugins which are freely distributed online were selected. The parameters used in the augmentation procedure were the ones set in the factory default preset for each audio plugin, except for those whose default preset did not alter significantly the sound. 

The audio effects used were the following:
\begin{itemize} 
 \item \textbf{Heavy distortion:} A Bitcrusher audio effect which produces distortion through the reduction of the sampling rate and the bit depth of the input sound was used in the training set. The VST plugin used for augmenting the training set was the TAL-Bitcrusher\footnote{\label{tal-fx} \url{https://tal-software.com/products/tal-effects}}. For the test and validation set, we used Camel Audio's CamelCrusher\footnote{\url{https://www.kvraudio.com/product/camelcrusher-}} plugin which provides distortion using tube overdrive emulation combined with a compressor.
 \item \textbf{Saturation:} For this effect, tube saturation and amplifier simulation plugins were used. The audio effect creates harmonics in the signal, replicating the saturation effect from a valve- or vacuum-tube amplifier \cite{zolzer2011dafx}. For this augmentation we focused on a subtle saturation which did not create noticeable distortion. The plugin used in the training set was the TAL-Tube\textsuperscript{\ref{tal-fx}}, while for the validation and test set Shattered Glass Audio's Ace\footnote{\url{http://www.shatteredglassaudio.com/product/103}} replica of a 1950s all tube amplifier was used.
 \item \textbf{Reverb:} To create a reverberation effect, the TAL-Reverb-4 plugin\footnote{\url{https://tal-software.com/products/tal-reverb-4}} was used in the test set. This effect replicates the artificial reverb obtained in a plate reverb unit. For the validation and test set we used OrilRiver\footnote{\url{https://www.kvraudio.com/product/orilriver-by-denis-tihanov}} algorithmic reverb, which models the reverb provided by room acoustics. The default preset for this plugin mimics the reverb present in a small room.
 \item \textbf{Echo:} A delay effect with long decay and with a big delay time (more than 50ms) \cite{zolzer2011dafx} was used to create an echo effect. We used the TAL-Dub-2\footnote{\url{ https://tal-software.com/products/tal-dub}} VST plugin in the training set and soundhack's ++delay\footnote{\url{ http://www.soundhack.com/freeware/}} validation and test set. For this last plugin, we adapted the factory default preset, changing the delay time to $181.7$\,ms and the feedback parameter to $50$\%, so that the echo effect was more noticeable.
 \item \textbf{Flanger:} For this delay effect, the input audio is summed with a delayed version of it, creating a comb filter effect. The time of the delay is short (less than $15$\,ms) and is varied with a low frequency oscillator \cite{zolzer2011dafx, reiss2014audio}. Flanger effects can also have a feedback parameter, where the output of the delay line is routed back to its input. For the training set, the VST plugin used was the TAL-Flanger\textsuperscript{\ref{tal-fx}}, while for the test and validation sets we used Blue Cat's Flanger\footnote{\url{https://www.bluecataudio.com/Products/Product\_Flanger/}}, which mimics a vintage flanger effect.
  \item \textbf{Chorus:} The chorus effect simulates the timing and pitch variations present when several individual sounds with similar pitch and timbre play in unison \cite{reiss2014audio}. The implementation of this effect is similar to the flanger. The chorus uses longer delay times (around $30$\,ms), a larger number of voices (more than one) and normally does not contain the feedback parameter \cite{zolzer2011dafx, reiss2014audio}. The VST effect used in the training set was the TAL-Chorus-LX\footnote{\url{https://tal-software.com/products/tal-chorus-lx}} which tries to emulate the chorus module present in the Juno 60 synthesizer. For the test and validation sets, we used Blue Cat's Chorus\footnote{\url{ https://www.bluecataudio.com/Products/Product\_Chorus}}, which replicates a single voice vintage chorus effect.
 \item \textbf{Pitch shifting:} For this effect, the LibROSA Python package for musical and audio analysis was used. This library contains a function which pitch shifts the input audio. As the dataset used contains recordings of the instruments for every note in the chromatic scale in successive octaves, our approach focused on pitch-shifting in steps smaller than one semitone, similarly to what can occur in a detuned instrument. The \texttt{bins\_per\_octave} parameter of the pitch-shifting function was set to  $72 = 12 \times 6$ while the \texttt{n\_steps} parameter was set to a random value between $1$ and $5$ for each sound. Neither $0$ or $6$ were selected as possible values as it would be the same as not altering the sound or pitch-shifting it by one semitone. The intention of the random assignment in the \texttt{n\_steps} is to ensure the size of this augmented dataset is equal to the size of the datasets of other effects.

\end{itemize}

The audio resulting from this augmentation step can be longer than the original unprocessed audio. In order to keep all examples with the same length, the processed audio files were trimmed, ensuring all audio samples had a fixed duration of $4$\,s, similar to the sounds presented in the NSynth dataset\cite{nsynth2017}.

The next step in the data processing pipeline is representing each sound in a log-scaled mel-spectogram. First, a $1024$-point Short-time Fourier transform (STFT) is calculated on the signal, with a $75$\% overlap. The magnitude of the STFT result is converted to a mel-spectogram with $80$ components, covering a frequency range from $40$\,Hz to $7600$\,Hz. Finally, the logarithm of the mel-spectogram is calculated, resulting in a $80\times247$ log-scaled mel-spectogram for the $4$\,s sounds sampled at $16$\,kHz present in the NSynth dataset \cite{nsynth2017}.

\subsection{Convolutional Neural Network}

The CNN architecture we chose to use in our experiment is the \textit{single-layer} architecture proposed by Pons et al.~\cite{pons2017timbre} for the musical instrument classification experiment, which has an implementation available online\footnote{\url{https://github.com/Veleslavia/EUSIPCO2017}}. This architecture uses vertical convolution filters in order to better model timbral characteristics present in the spectogram, achieving close to state-of-the-art results \cite{classificationHan}, using a much smaller model (23 times less trainable parameters) and a consequently lower training time.

We chose the \textit{single-layer} architecture presented in this study and adapted it to take an input of size $80\times247$. This architecture contains a single but wide convolutional layer with different filters with various sizes, to capture the timbral characteristics of the input: 
\begin{itemize}
 \item $128$ filters of size $5\times1$ and $8\times1$;
 \item $64$ filters of size $5\times3$ and $80\times3$;
 \item $32$ filters of size $5\times5$ and $80\times5$.
\end{itemize}
\begin{figure*}[h]
 \includegraphics[width=\textwidth]{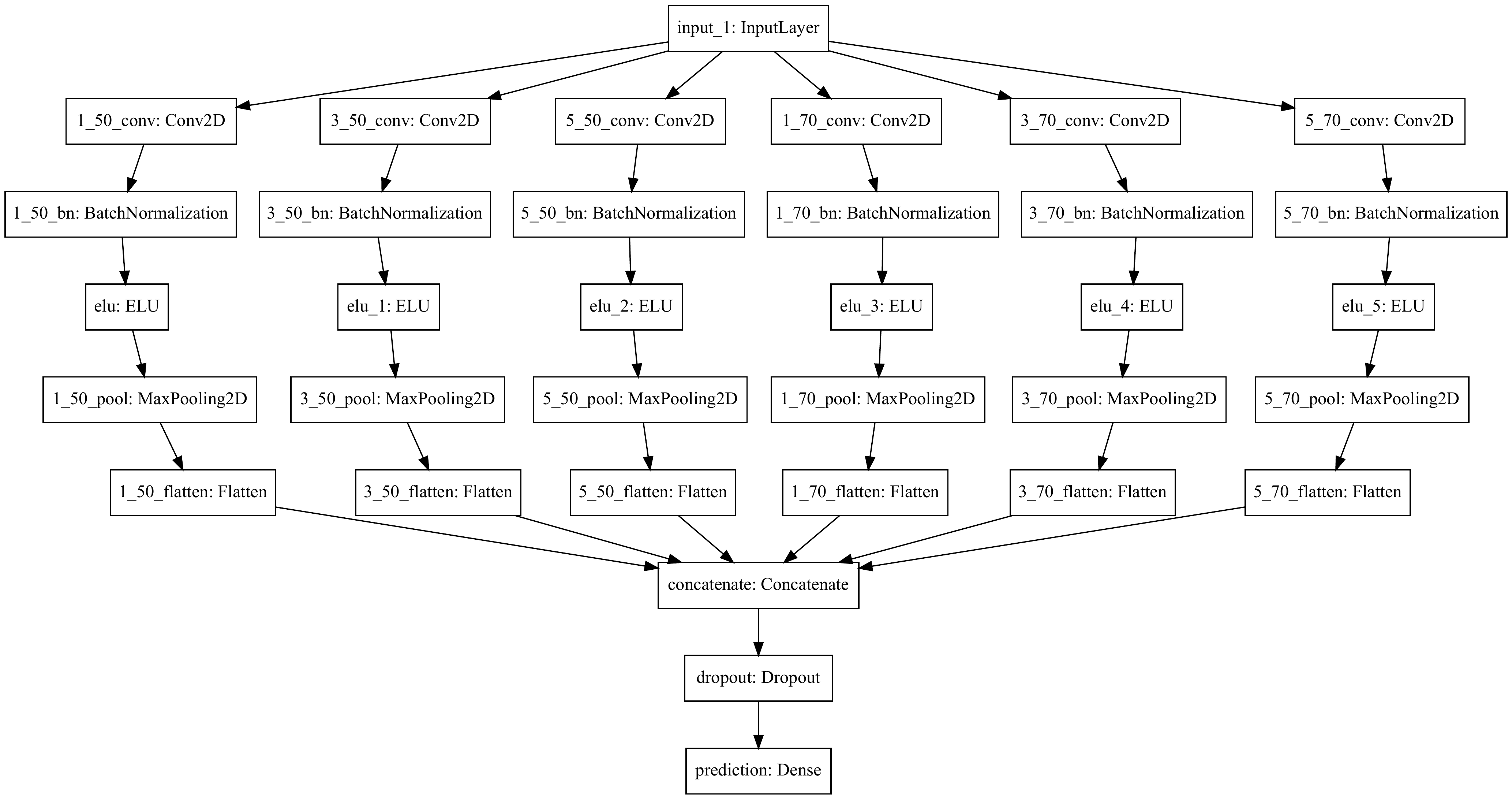}
 \caption{Single-layer CNN architecture proposed in \cite{pons2017timbre}}
 \label{fig:1}
\end{figure*}

Batch normalisation \cite{ioffe2015batch} is used after the convolutional layer and the activation function used is Exponential Linear Unit \cite{elu}. Max pooling is applied in the channel dimension for learning pitch invariant representations. Finally, $50$\% dropout is applied to the output layer, which is a densely connected $11$-way layer, with the \textit{softmax} activation function. A graph of the model can be seen in Figure \ref{fig:1}. For more information on this architecture and its properties see \cite{pons2017timbre}.

\subsection{Evaluation}

The training of the models used the Adam optimiser \cite{kingma2014adam}, with a learning rate of $0.001$. In the original paper \cite{pons2017timbre} the authors used Stochastic Gradient Descent (SGD) with a learning rate reduction every $5$ epochs. This was shown to provide good accuracy on the IRMAS dataset. However, we chose to use Adam as an optimiser because it does not need significant tuning as SGD. Furthermore, using a variable learning rate dependent on the number of epochs could benefit the larger training datasets as is the case of the ones with augmentation. A batch size of $50$ examples was used, as it was the largest batch size able to fit the memory of the available GPUs. The loss function employed for the training was the categorical cross-entropy, as used in \cite{pons2017timbre}, which can be calculated as shown in Equation (\ref{eq:1}), where $N$ represents the number of observations (examples in the training set) and $p_{model}[y_i \in C_{y_i}]$ is the predicted probability of the $i^{th}$ observation belonging to the correct class $C_{y_i}$. 

\begin{equation} \label{eq:1}
loss = -\frac{1}{N}\sum_{i=1}^{N}\log p_{model}[y_i \in C_{y_i}]
\end{equation}

To compare the models trained with the different datasets, we used categorical accuracy as evaluation metric, described in Equation (\ref{eq:2}). A prediction is considered correct if the index of the output node with highest value is the same as the correct label.
\begin{equation} \label{eq:2}
    \text{Categorical Accuracy} = \text{Correct predictions} / \text{N}
\end{equation}
All the models were trained until the categorical accuracy did not improve in the validation set after $10$ epochs and the model which provided the best value for the validation set was evaluated in the test set.

\section{Results}

Two experiments were conducted in our study. We firstly evaluated how augmenting the training set of NSynth \cite{nsynth2017} by applying audio effects to the sounds can improve the automatic classification on the instruments of the unmodified test set. In the second experiment we evaluated how robust a state of the art model for instrument classification is when classifying sounds where these audio effects are applied.

\begin{table}[]
\centering
\caption{\label{tab:1}Classification accuracy on the unprocessed test set.} 
\begin{tabular}{ccc}

\hline
Test Effect           & Train Effect   & Accuracy \\ \midrule
\multirow{8}{*}{None} & None (baseline) & 0.7378   \\
                      & Heavy distortion     & \textbf{0.7473}   \\
                      & Saturation     & 0.7349   \\
                      & Reverb         & 0.7375   \\
                      & Chorus         & \textbf{0.7417}   \\
                      & Echo           & 0.7336   \\
                      & Flanger        & \textbf{0.7412}   \\
                      & Pitch Shifting & 0.7334 \\ \bottomrule
\end{tabular}
\end{table}

The results of the first experiment are presented in Table \ref{tab:1}, where the classification accuracy between the models trained with the original NSynth training set augmented with audio effects can be compared to the baseline (unprocessed dataset). We see that the increase in accuracy only occurs for chorus, heavy distortion and flanger effects. The highest classification accuracy was achieved by the dataset augmented with heavy distortion, where an increase of 1\% was obtained. However, all the accuracy values are in a small interval (between 0.7334 and 0.7473), which means that the model was not able to learn from the augmented datasets. Future experiments are needed in order to understand why this occurs. In \cite{salamon2017deep}, the authors state that the superior performance obtained was due to an augmentation procedure coupled with an increase in the model capacity. Experiments with higher capacity models will be performed to understand if the size of the model used is limiting its performance on learning from the augmented dataset.

\begin{table}[]
\caption{\label{tab:2}Classification accuracy on the augmented test set.}
\centering
\begin{tabular}{ccc}
\toprule
Test Effect    & Train Effect   & Accuracy \\ \midrule
Heavy distortion     & None           & 0.3145   \\
               & Heavy distortion     & \textbf{0.3518}       \\ \midrule
Saturation     & None           & \textbf{0.4836}   \\
               & Saturation     & 0.4607   \\ \midrule
Reverb         & None           & \textbf{0.3931}   \\
               & Reverb         & 0.3774   \\ \midrule
Chorus         & None           & 0.6348   \\
               & Chorus         & \textbf{0.6436}   \\ \midrule
Echo           & None           & \textbf{0.4719}   \\
               & Echo           & 0.4319   \\ \midrule
Flanger        & None           & \textbf{0.7046}   \\
               & Flanger        & 0.7002   \\ \midrule
Pitch Shifting & None           & \textbf{0.6980}   \\
               & Pitch Shifting & 0.6741  \\ \bottomrule
\end{tabular}
\end{table}

In Table \ref{tab:2}, we present the accuracy values obtained when evaluating the trained model on test sets processed with effects. The first thing we verify is that the accuracy of the classification greatly decreases for almost all effects, when compared to the unprocessed sound classification. The model seems to be more robust to the flanger and to the pitch shifting effect, where the difference between the accuracy on the unprocessed test set and on the processed one is smaller than 4\%. The effects which caused the biggest drops in accuracy ( > 20\% ) were the heavy distortion, the saturation, the echo and the reverb. When evaluating if training with the augmented datasets increased the robustness of the model, we see that this is only true for the chorus and distortion effect. While for the heavy distortion effect the accuracy when training with the augmented set is improved by a significant value ($\approx 4\%$), the difference in accuracy between training with the augmented and the unprocessed sets are small. Further experiments will be performed to understand the bad generalisation of the model. Besides experimenting with a higher capacity model as previously stated, work will be conducted on further augmenting the datasets. Although the effects applied were the same in the training, validation and test sets, the implementations used were different in the training set. This leads to a different timbre between the sets that the architecture might not be able to generalise to. In future experiments, we will further augment the dataset using a number of different settings for each effect, as well as different combinations of the effects applied.


\section{Conclusions}

In this paper we evaluated how a state of the art algorithm for automatic instrument classification performs when classifying the NSynth dataset and how augmenting this dataset with audio effects commonly used in electronic music production influences its accuracy on both the original and processed versions of the audio. We identify that the accuracy of this algorithm is greatly decreased when tested on sounds where audio effects are applied and see that the augmentation can lead to better classification in unprocessed sounds. We note that the accuracy results provided are preliminary, and do not fully exploit the possibilities of using audio effects for data augmentation in automatic instrument classification. We are currently evaluating how a deeper architecture performs on the same task. Further work includes evaluating how using a bigger variety of effects, with different combinations of parameters, further improves the robustness of the classification algorithm.

\section{Acknowledgments}
This project has received funding from the European Union's Horizon 2020 research and innovation programme under the Marie Sk\l{}odowska-Curie grant agreement N\textsuperscript{o} 765068, MIP-Frontiers. We thank Matthew Davies for reviewing a draft of this paper and providing helpful feedback.
\nocite{*}
\bibliographystyle{IEEEbib}
\bibliography{DAFx19_tmpl} 

\end{document}